# Vibrational Entropy Investigation in High Entropy Alloys

## Abstract

Mina Aziziha[1], Saeed Akbarshahi[1]

[1]Department of Physics & Astronomy, West Virginia University, Morgantown, WV 26506, USA

The lattice dynamics for NiCo, NiFe, NiFeCo, NiFeCoCr, and NiFeCoCrMn medium to high entropy alloy have been investigated using the DFT calculation. The phonon dispersions along three different symmetry directions are calculated by the weighted dynamical matrix (WDM) approach and compared with the supercell approach and inelastic neutron scattering. We could correctly predict the trend of increasing of the vibrational entropy by adding the alloys and the highest vibrational entropy in NiFeCoCrMn high entropy alloy by WDM approach. The averaged first nearest neighbor (1NN) force constants between various pairs of atoms in these intermetallic are obtained from the WDM approach. The results are discussed based on the analysis of these data.

## Introduction

Alloys attracted a lot of attention because of their unique ability to change the physical properties of materials[1–4]. When there are several components (5 or more) participating in the alloying process, the resulting single-phase solid-solution alloy might be called high entropy alloy (HEA) if the high configurational entropy which is found responsible for the stability of the single solid solution phase over competing intermetallic and elemental phases. However, there are several other contributions in the Gibbs free energy, which determines phase stabilities. The aim of the new concept in the development of new alloy systems with novel mechanical and functional properties [5,6]. Therefore, the role of entropy contributions investigation is vital. Many types of single-phase HEAs have been developed with different crystal structures such as face-centered cubic (FCC) [7], body-centered cubic (BCC), and hexagonal close-packed (HCP) [8].

In this work, we employ ab initio-based simulations to investigate the vibrational entropy which is one of the entropy contributions (such as vibrational, electronic, magnetic as well as configurational) for NiCo, NiFe, NiFeCo, NiFeCoCr, and NiFeCoCrMn. NiCo, NiFe, and NiFeCo. All of the alloys are made of transition metals, and they ave strong force-constant disorder but minimal mass disorder. The contributing elements are isoelectronic and selected from the same row of the periodic table (either 3*d* or 4*d*) and have a similar atomic size. We made a comparison between WDM, Supercell approach, and experimental results on the phonon dispersion. Then, we predicted the maximum vibrational entropy for NiFeCo, NiFeCoCrMn.

**Computational Details**

The density functional theory is used for performing the calculations [9,10] with a plane-wave basis set, as implemented in the Quantum-Espresso [11]. We used the Perdew–Burke–Ernzerhof generalized gradient approximation exchange-correlation functional [12,13] and Optimized Norm-Conserving Vanderbilt Pseudopotential [14]. The relaxation is done for a variable cell-structure in Quantum-Espresso until the Hellmann-Feynman force and stress reach to less than 1mRy/Bohr and 0.1 mRy/Bohr [15,16]. The relaxed primitive unit cell with cubic structure ($Fm\bar{3}m$) (225) of Ni, Fe, Co, Cr, and Mn is used to construct a 4×4×4 supercell of 64 atoms. Also, Spin-polarization is considered because the elements are magnetic. The relaxation of this supercell is done with a $6 \times 6 \times 6$ Monkhorst-Pack *k*-point grid [17]. The energy cut-off 110 Ry for wave functions was employed for calculations. To calculate the force constants using PHONOPY [18] software. In the supercell method for alloys, a disordered finite-size supercell with defects that break the space group symmetry of the crystal leads to a shrinking Brilluion zone in reciprocal space. The band unfolding methods have been developed for electronic problems to recover the phonon spectra within the Brillouin zone of the primitive cell [1,19–21]. Here, to unfold the phonon band, we use the program developed by Ikeda et al. For supercell calculation of the NiFe, NiCo, NiFeCo, we used the data (108-atom supercell model) reported by Sai Mu et al.[22]

**Weighted Dynamical Matrix (WDM) Approach**

To find the phonon modes, one needs to construct the dynamical matrix. To do so, $r_{li}$ The deviation of the $i^{th}$ atom in the $l^{th}$ unit cell of the Supercell from equilibrium position is introduced. Also, the second-order force constants for the $li - l'i'$ atom pair denoted as $\Phi_{\alpha\beta}(0i, l'i')$ (with $\alpha$ and $\beta$ indices for coordinates) are calculated. Then, the dynamical matrix **D(q)** at the wavevector **q** is constructed as follows:

$$D_{ii'}^{\alpha\beta}(\mathbf{q}) = \frac{1}{\sqrt{m_i m_{i'}}} \sum_{l'} \Phi_{\alpha\beta}(0i, l'i') \exp[i\mathbf{q} \cdot (\mathbf{r}_{l'i'} - \mathbf{r}_{0i})], \tag{1}$$

where $m_i$ is the mass of the $i^{th}$ atom. Phonon frequencies $\omega(\mathbf{q}, \kappa)$ and mode eigenvectors $\boldsymbol{\chi}(\mathbf{q}, \kappa)$ at **q**, where $\kappa$ is the band index, are obtained by solving the eigenvalue equation:

$$\mathbf{D}(\mathbf{q})\,\boldsymbol{\chi}(\mathbf{q}, \kappa) = [\omega(\mathbf{q}, \kappa)]^2 \boldsymbol{\chi}(\mathbf{q}, \kappa). \tag{2}$$

To calculate the phonon modes for these alloy samples, first, the Hellmann-Feynman forces of the parent structures (Cu and Au), for the atomic displacement of 0.01 Å from their equilibrium positions, are calculated by DFT calculations. Then from forces $\vec{F}_{Cu}$ and $\vec{F}_{Au}$ the force constants $\Phi_{Cu}$ and $\Phi_{Au}$ are calculated. The weighted dynamical matrix is constructed as follows:

$$\overline{D}_{ii'}^{\alpha\beta}(\mathbf{q}) = \frac{1}{\sqrt{m_i m_{i'}}} \sum_{l'} \overline{\Phi}_{\alpha\beta}(0i, l'i') \exp[i\mathbf{q} \cdot (\mathbf{r}_{l'i'} - \mathbf{r}_{0i})], \tag{3}$$

See more details on the WDM approach in references [23–25].

**Results and Discussion**

We used the WDM approach to calculate the vibrational properties for NiCo, NiFe, and NiFeCo, which are shown in Figure 1 (a)-(c). The phonon dispersion curves and phonon density of states results are obtained along [0,0,ζ], [0,ζ,ζ] and [ζ,ζ,ζ] directions. The WDM calculations and the experimental results of phonon frequencies agree reasonably well on transverse and longitude modes along [0,0,ζ], [0,ζ,ζ] and [ζ,ζ,ζ] directions ($\zeta = \frac{\vec{q}}{\vec{q}_{max}}$, $\vec{q}$ is the phonon wave-vector). The lighter atoms correspond to the higher frequency regions, and heavier atoms correspond to the lower frequency regions. We compared our results to the inelastic neutron scattering and supercell calculation results reported by Sai Mu et al. [22]. We calculated the phonon dispersion by WDM for the NiFeCoCr and NiFeCoCrMn. The results are plotted in Figure 2 (a) and (b). The disordered supercell calculation is a computationally heavy calculation, but using the WDM approach, the computational load decreased significantly.

Change of the vibrational entropy with increasing the number of elements is important for phase stability considerations. To test whether the results can support the previous studies, we calculated the vibrational entropy for NiFe, NiFeCo, NiFeCoCr, and NiFeCoCrMn, which are plotted in Figure 4. The results are showing a good agreement with the supercell computed result shown in red color and for the NiFeCoCrMn with a solid blue circle. Furthermore, in Figure 5 the vibrational entropy is calculated for CoCr, CoCrFe, CoCrFeMn and CoCrFeMnNi. We can see when alloying NiFeCo as a medium entropy with Cr, vibrational entropy significantly increases. Because the mass of the Cr is lighter than Ni, Fe, and Co elements. Also, an increasing trend in the vibrational entropy is observed with an increase in the average mass which are 57.83, 56.37, and 56.08 (AMU), respectively, ($M_{NiFeCo} > M_{NiFeCoCr} > M_{NiFeCoCrMn}$). The highest predicted vibrational entropy has an average mass $M_{Ni_5Fe_5Co_5Cr_{30}Mn_{55}} = 54.49$ (AMU). Figure 3 (a) to (d) shows the change of entropy ternary diagram for NiFeCo, NiCoCr, CoCrFe, and NiFeCoCrMn, respectively, versus concentration. Visualization is performed using the MPLTERN code in combination with MATPLOTLIB [26]. The highest entropy that can be achieved for each combination is marked by the star symbol. The highest entropy is observed for the $Ni_5Fe_5Co_{90}$, $Ni_5Co_{65}Cr_{30}$, $Co_{65}Cr_{30}Fe_5$, and $Ni_5Fe_5Co_5Cr_{30}Mn_{55}$ with 8.18, 8.47, 8.47 $k_B$, and 8.53 $k_B$, respectively.

The first nearest-neighbor force constants, which shown in Table 1, are an order of magnitude larger than those of the further neighbors, so we used them to comparison our results and the behavior of force constant versus bond distance follows the expected trend

**Conclusions**

We have investigated the lattice dynamics NiCo, NiFe, NiFeCo, NiFeCoCr, and NiFeCoCrMn medium to high entropy alloys using first-principle density functional theory. The agreement between WDM, supercell approach, and neutron scattering experiments is good, but our results overestimate frequencies for NiFe. That might be due to the lack of exchange-correlation function and averaging of force-constant. This is important from the point of view of the feasibility of using *ab initio* ordered alloy force constants to study the disordered high entropy alloys and correctly predict the highest vibrational entropy of them, which are $Ni_5Fe_5Co_{90}$ and $Ni_5Fe_5Co_5Cr_{30}M.n_{55}$. We observe the trend of increasing the vibrational entropy with an increase of the average mass, and the results are analyzed using the nearest-neighbor force constant values between various pairs of species. The behavior of force constant vs. bond distance follows the expected trend.

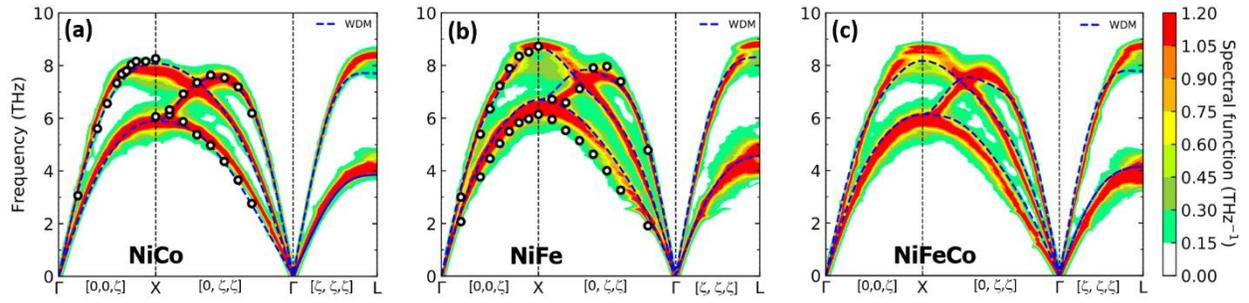

Fig. 1: (a)-(c) Comparison of phonon dispersion of the NiCo, NiFe, and NiFeCo along $[0, 0, \zeta]$, $[0, \zeta, \zeta]$ and $[\zeta, \zeta, \zeta]$ directions between WDM and Supercell approach. The blue dashed line in the figures represent the WDM calculations, the circle in the figures represent the experimental results.

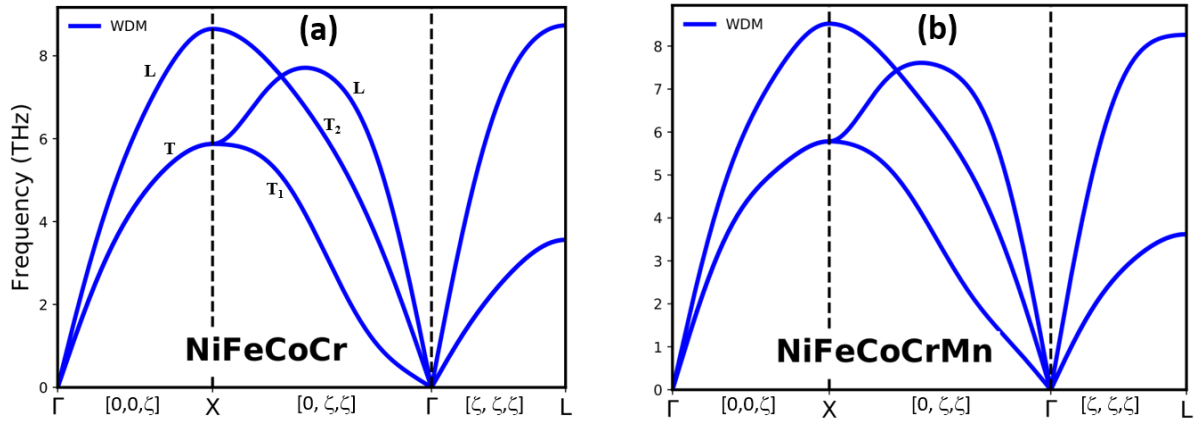

Fig. 2: (a)-(b) Phonon dispersion of NiFeCoCr and NiFeCpCrMn along $[0, 0, \zeta]$, $[0, \zeta, \zeta]$ and $[\zeta, \zeta, \zeta]$ directions by WDM approach. The blue line in the figures represent the WDM calculations.

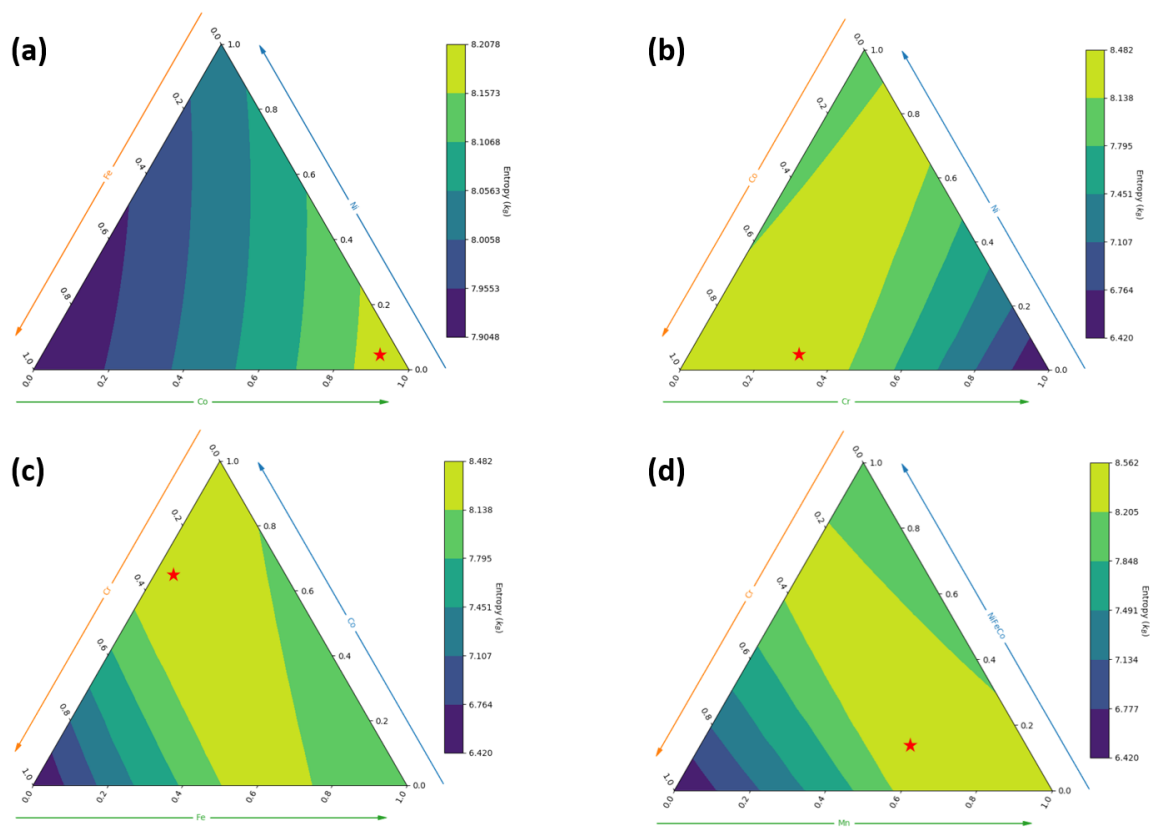

Fig. 3: Impact of alloying on the entropy. (a)-(c) change of entropy of the NiFeCo, NiCoCr and CoCrFe versus concentration. The highest entropy of the configures are $Ni_5Fe_5Co_{90}$, $Ni_5Co_{65}Cr_{30}$ and $Co_{65}Cr_{30}Fe_5$ with 8.18, 8.47 and 8.47 $k_B$ respectively (d) change of entropy of the NiFeCo versus Cr and Mn. The red star represents the configuration with largest entropy $Ni_5Fe_5Co_5Cr_{30}Mn_{55}$ with amount 8.53 $k_B$.

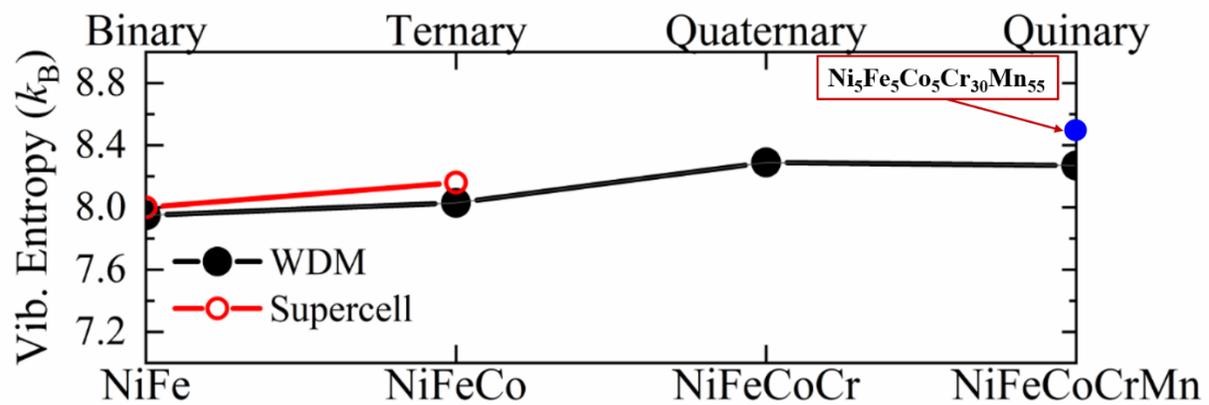

Fig. 4: Vibrational entropy for equiatomic NiFe, NiFeCo, NiFeCoCr and NiFeCoCrMn at 1500 K. The black solid circle is WDM calculation results. The red line is Supercell calculation results.

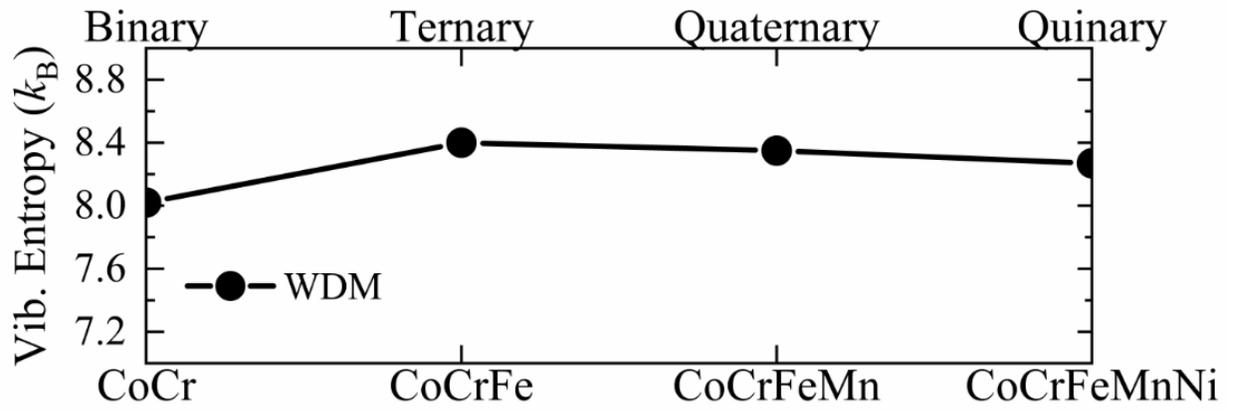

Fig. 5: Vibrational entropy for equiatomic CoCr, CoCrFe, CoCrFeMn and CoCrFeMnNi at 1500 K. The black solid circle is WDM calculation results. The red line is Supercell calculation results.

Table 1 The averaged first Nearest Neighbor (1NN) force constants, lattice parameter and bond length in Cr, Ni, Co, Fe, Mn, NiFe, NiCo extracted from supercell calculations. The force constants, lattice parameter, band length and mass are given in eV/Å$^2$, Å, Å and AMU respectively.

|  | Average 1NN $\Phi_{xx,yy,zz}$ (eV/Å$^2$) | Lattice parameter (Å) | Bond length (Å) | Mass (AMU) |
|---|---|---|---|---|
| Cr-Cr | -0.809 | 3.61 | 2.55 | 51.996 |
| Ni-Ni | -0.734 | 3.50 | 2.48 | 58.693 |
| Co-Co | -0.520 | 3.51 | 2.48 | 58.933 |
| Fe-Fe | -0.622 | 3.46 | 2.44 | 55.845 |
| Mn-Mn | -0.319 | 3.50 | 2.48 | 54.938 |
| NiFe | -0.678 | - | - | 57.270 |
| NiCo | -0.627 | - | - | 58.813 |
| NiFeCo |  |  |  | 57.830 |
| NiFeCoCr |  |  |  | 56.370 |
| NiFeCoCrMn |  |  |  | 56.080 |